# Quantum simulation of a spin polarization device in an electron microscope

**Vincenzo Grillo[1,2], Lorenzo Marrucci[3,4], Ebrahim Karimi[5], Riccardo Zanella[6], Enrico Santamato[3]**

1 CNR-Istituto Nanoscienze, Centro S3, Via G Campi 213/a, I-41125 Modena, Italy
2 CNR-IMEM, Parco delle Scienze 37a, I-43100 Parma, Italy
3 Dipartimento di Fisica, Università di Napoli ''Federico II'', Italy
4 CNR-SPIN, Complesso Universitario di Monte S. Angelo, Napoli, Italy
5 Department of Physics, University of Ottawa, Ottawa ON K1N 6N5  Canada
6 Laboratory for Technologies of Advanced Therapies (LTTA), Via Fossato di Mortara 70, 44121 Ferrara

E.mail: vincenzo.grillo@unimore.it

**Abstract.** A proposal for an electron-beam device that can act as an efficient spin-polarization filter has been recently put forward [E. Karimi et al., Phys. Rev. Lett. **108**, 044801 (2012)]. It is based on combining the recently developed diffraction technology for imposing orbital angular momentum to the beam with a multipolar Wien filter inducing a sort of artificial non-relativistic spin-orbit coupling. Here we reconsider the proposed device with a fully quantum-mechanical simulation of the electron beam propagation, based on the well established multi-slice method, supplemented with a Pauli term for taking into account the spin degree of freedom. Using this upgraded numerical tool, we study the feasibility and practical limitations of the proposed method for spin-polarizing a free electron beam.

## 1. Introduction

In electron microscopy, the electron spin has been, for a long time, a spectator degree of freedom. The main reason is that it is very hard to produce good quality spin-polarized electron beams. Indeed, at the present state of the art of electron microscopy, this can only be done by exploiting laser-induced emission from a strained semiconductor [ 1][ 2], a process that cannot reach a very high brightness and polarization purity at the same time, and that typically requires the frequent replacement of the emitting tip. The introduction of a spin polarizer for free electrons hence would be a breakthrough in the generation of spin-polarized electron beams with high brilliance.

The question of if a device capable of spin-polarizing free electron beams could ever exist has been largely debated in the scientific literature. The first attempts date back even to the early times of quantum mechanics. The failure of the first experiments and the debate on their possible improvements led to the idea that fundamental reasons hamper the realization of such a polarizer [ 3]. More explicitly,  Bohr's assertion, as formulated by Pauli, was that  "it is impossible to observe the spin of the electron, separated fully from its orbital momentum, by means of experiments based on the concept of classical particle trajectories" [ 4].

This sort of interdiction discouraged until recently the attempts to create a spin polarizer for an electron beam based on classic trajectory manipulation. The reason for the denial of these experiments was always tied down, more or less directly, to the uncertainty principle [ 4]. Among the proposed experiments [ 5], there was a variant of the famous experiment of Stern and Gerlach, in which a gradient of the magnetic field was exploited to separate the trajectories of particles with different spins [ 6]. The refutation, in this case, was based on the argument that an additional Lorenz force must exist because of the ineliminable component of the magnetic field perpendicular to its gradient. This led to the proposal by Knauer [ 7], again based on the Stern Gerlach concept, but with the addition of an electric field for compensating the Lorentz force. This idea was also proved to be ineffective by Bohr and Pauli, owing to the unavoidable effect of the fringe fields on the electrons entering the device. Looking at the details of Bohr's interdiction argument, one may notice that it does not deny the possibility of spin polarization in general, but only of methods based on classical physics concepts. In this sense, the prediction and following discovery of the partial polarization of electrons undergoing Mott scattering cannot be considered as a violation of Bohr's statement [8][9]. Indeed, the crucial factor determining the polarization in Mott scattering is the spin-orbit potential, which has no direct classic analogue. Moreover, a recent reanalysis of "anti-Bohr" experiments has actually shown that even in the classic Stern-Gerlach-like experiments a small but non-vanishing polarization can be obtained [10].

In this context, we have proposed an experiment based on the use of electron vortex beams, i.e. beams carrying orbital angular momentum (OAM)[11]. We took advantage of the fact that electron vortex beams have been recently demonstrated experimentally [12][13][14].

The advantage of such beams is that the OAM can be coupled with the spin in the interaction with typical electron-optical elements such as the magnetic quadrupoles used for beam stigmatization. The most promising implementation of the proposed polarization experiment is based on a quadrupolar magnetic field combined with an electric quadrupole, in order to cancel the Lorentz force. Such device is sometimes called a second-order Wien filter.

To a first view, this proposed experiment looks similar to that of Knauer [7]. The main difference apparently is in the shape of the magnetic (and electric) field. In the new method the field pattern is characterized by the presence of a topological charge. The latter implies the presence of an ineliminable singularity related to the topology of the field, as for example in the case of the quadrupolar field, which goes to zero (and therefore has singular phase) in its center. Explicitly the topological charge q enters in the analytical description of the magnetic field, which can be expressed as

$$\overline{B} = B_0(r) \left[\cos(q\theta + \beta), \sin(q\theta + \beta), 0\right] \tag{1}$$

where $r$, $\theta$ are the polar coordinates in the plane transverse to the beam propagation axis z, $\beta$ is a real constant, and the field radial profile is $B_0(r) \propto r^{-q}$. The topological charge q is an integer, which must be negative if we assume that there is no field source at the element center, r = 0. The simplest case is the quadrupole, for which q = –1. We introduced the name "q-filter" to denote a higher-order Wien filter with fields having the geometry given by Eq. (1).

Owing to this topological charge, a non vanishing Berry phase is associated to the adiabatic transport around the singularity of the spin precession solution. In fact, while the classical spin precesses in presence of a magnetic field, a different phase is associated to each "trajectory" inside the quadrupole. This phase, in turn, gives rise to a spin-dependent quantum interference effect in the radial direction, allowing for the electron separation according to their spin along this coordinate. Despite its superficial similarity to Knauer's proposal, our method, being based on a quantum interference, is again clearly non-classical, and hence it can escape Bohr's interdiction.

In our previous work, only a semi-classical analysis of the electron propagation in the filter has been carried out [11]. However, the subtle arguments used against the Stern-Gerlach experiment prove that limiting factors can be easily overseen in any qualitative or approximate analysis. A fully quantum simulation of the electron wavefunction propagation is therefore in order. To this aim, in this work, we introduce a simulation algorithm derived from the multi-slice method used for simulating electron-matter interaction, which is one of the most powerful tools to calculate dynamical scattering in transmission electron microscopy (TEM) [15]. The multi-slice simulation, in its basic form, is based on paraxial high-energy forward scattering and is typically applied to the interaction of TEM electron with crystals. Nonetheless, it has been demonstrated that the interaction of the electromagnetic lenses of a microscope with the beam can also be satisfactorily accounted for by this algorithm [16].

The advantage of the multi-slice algorithm is that it retains the wave description of the electrons. In many cases, such as in the propagation of electrons through electromagnetic lenses, this approach may result time-consuming and not truly advantageous with respect to classical ray-tracing. But it becomes essential in the description of purely quantum effects, such as the Aharonov-Bohm problem [17][18]. In the present case, the use of a quantum-mechanical approach is necessary to capture the role of the spin in the wavefunction propagation in the presence of spin-orbit effects [19].

Previous attempts to produce a spin-dependent multi-slice approach to describe electron-matter interaction have been based on the relativistic description of electron propagation derived from the Dirac equation [20]. In the present case we limit ourselves to the non-relativistic case, adding the sole Pauli term in the interaction Hamiltonian, while postpone to future work a full evaluation of all the relativistic corrections.

Based on this approach, we aim to reconsider in greater detail the proposed procedure to create beams of polarized electrons. Particular attention will be given to the polarized intensity in relation to the applied field, to the role of the fringing fields, and to the required emission coherence.

## 2. Non-relativistic simulation model

As the main optical element in the calculation is the Wien filter, it is necessary to describe the effect of both electrostatic and magnetic fields on a free electron. The non-relativistic Pauli Hamiltonian for the electron dynamics can be written as follows:

$$H = \frac{(-i\hbar\bar{\nabla}+e\bar{A})^2}{2m} - eV_E - \bar{B}\cdot\bar{\mu} \tag{2}$$

where $\hbar$ is the reduced Plank constant, m is the electron rest mass, $e$ is the (absolute value of the) electron charge, $\bar{A}$ is the magnetic vector potential, $V_E$ is the electrostatic potential, and $\bar{\mu} = -g\frac{e\hbar}{2m}\frac{\bar{\sigma}}{2}$ is the magnetic momentum vector, with the factor g≈2 being the gyromagnetic g-factor, and $\bar{\sigma}$ the Pauli matrix vector.

This Hamiltonian can be split into three terms that in the Coulomb gauge are as follows:

$$H = H_0 + U + U_p \tag{3}$$

where

$$H_0 = -\frac{\hbar^2\nabla^2}{2m} \tag{4a}$$

$$U = -eV_E + \frac{e^2}{2m}A^2 - \frac{\hbar e}{m}i\bar{A}\cdot\bar{\nabla} \tag{4b}$$

$$U_p = \frac{g}{2}\frac{e\hbar}{2m}\bar{B}\cdot\bar{\sigma} \tag{4c}$$

The paraxial approximation permits to simplify the kinetic-energy hamiltonian $H_0$ as follows:

$$H_0 = -\frac{\hbar^2\nabla_T^2}{2m} - \frac{i}{m}\hbar^2 k_0 \frac{\partial}{\partial z} \tag{5}$$

where $\nabla_T$ is the gradient operator in the transverse plane *xy* and z is the main propagation direction. The quantity $k_0$ is the wave number, inversely proportional to the electron wavelength $\lambda$ and can be written

$$k_0 = \frac{2\pi}{\lambda} = \frac{1}{\hbar}\sqrt{2m\varepsilon\left(1+\frac{\varepsilon}{2mc^2}\right)} \tag{6}$$

where $\varepsilon$ is the kinetic energy imposed by the accelerating voltage and c is the seed of light in vacuum. It is appropriate to introduce here also the related quantities that describe the forward propagation and that are treated as constant throughout this article namely the average velocity v and the relativistic mass as a function of the kinetic energy $\varepsilon$

$$v = \frac{\hbar k_0}{m^*} = c\sqrt{1-\frac{1}{(1+\frac{\varepsilon}{mc^2})^2}} \tag{7}$$

$$m^* = \gamma m = m\left(1+\frac{\varepsilon}{mc^2}\right). \tag{8}$$

The U term of Eq. 4b gives the spin-independent interaction term with the applied electromagnetic field. If the magnetic field B is everywhere orthogonal to the propagation, as in the core of the Wien filter, we can write

$$U = -eV_E + \frac{e^2}{2m}A^2 - \frac{\hbar e}{m}iA_z\frac{\partial}{\partial z} \qquad (9)$$

The third term of U is usually the dominating one, responsible for the main Lorentz force effects and can be further simplified in the forward scattering approximation as $\frac{\hbar e}{m}A_z k_0$.

Conversely if the magnetic field is oriented along z, as, for example, in the center of the objective lens of a microscope, one has

$$U = -eV_E + \frac{e^2}{2m}A^2 - \frac{\hbar e}{m}i(A_x\frac{\partial}{\partial x} + A_y\frac{\partial}{\partial y}) \qquad (10)$$

The second term provides the lensing effect responsible for localization, while the third one produces, for an homogeneous field, a coordinate rotation [21].

Finally the $U_p$ term given in Eq. 4c, typically excluded in microscopy simulations, is the Pauli term, describing the interaction between the magnetic dipole of the electron and the external magnetic field.

**3. Main relativistic corrections**

In this section the main relativistic corrections to the electron motion, taking into account the spin, will be considered, following in part the approach of Rother [20].

The first somehow more obvious correction should be applied to the dispersion equation, namely in the term $H_0$. An expansion of the exact relativistic Hamiltonian in power of the momentum p would yield a series of correction terms, the first of which is [22]

$$H_0' = -\frac{P^4}{8m^3c^4} \qquad (11)$$

Since all these terms commute with $H_0$ and can be treated perturbatively, it is customary to include these effects in the substitution of the rest mass with its relativistic counterpart m* [22].

More complex to deal with are the terms that contain the charge and spin interaction with external fields. While the non-relativistic treatment of the spin requires a 2-component spinor, the full relativistic treatment derived from Dirac equation would require the introduction of a 4-component spinor. In fact, in the rest frame of the electron the two additional components vanish identically.

To correct our approach, we refer to the work of Rother et al. for a fully relativistic multi-slice method [20]. This article highlights the presence of two Hamiltonian correction terms not included so far

$$U_R = \frac{e^2}{2m}\frac{(V_E)^2}{c^2} + \frac{2m}{\hbar^2 c}\bar{\alpha}\cdot\bar{E} \qquad (12)$$

where $\bar{\alpha} = \begin{bmatrix} 0 & \bar{\sigma} \\ \bar{\sigma} & 0 \end{bmatrix}$ in the standard representation an $\bar{E}$ is the electric field. The second term is responsible for spin-orbit effects. We note in passing that the Darwin term vanishes here, since, in our case, $\bar{\nabla}\cdot\bar{E} = 0$.

These two terms are responsible for complex effects depending on the electric field. In the case of interest of a Wien filter, for which $\bar{E}$ and $\bar{B}$ are compensated, their role can be better clarified.

Since in this case $V_E = vA_z$ the term $V_E^2$ has a similar effect as the term $A^2$. Therefore a way to account for this in the simulation would be to substitute

$$A_{eff}^2 = A^2\left(1 + \frac{v^2}{c^2}\right) \qquad (13)$$

But we will not consider this correction. In fact, in agreement with what stated in the literature and anticipated above, we assume that the spin-independent electron spatial evolution can be satisfactorily described by the non-relativistic equations modified only with the relativistic corrections for mass and wavelength [22] (a more comprehensive discussion can be found in [23]).

What remains to be considered are therefore only the relativistic terms in the Pauli interaction. The term $\bar{\alpha} \cdot \bar{E}$ in Eq. (12) introduces an interaction between spin and electric field, but also involves the lower part of the quadrispinor that we are going to neglect. The size of this correction would once again go as $v^2/c^2$. However we prefer to calculate this term in a more "classical" way. This can be done by requiring the relativistic covariance of the forces acting on electron in presence of an electric field. In the reference frame at rest with the electron, the electric field can be seen as an additional magnetic field providing an additional $\bar{B} \cdot \bar{\mu}$ Hamiltonian. The first-order approximation of this effect, in the hypothesis of a perfectly compensated Wien filter, gives rise to a reduction of the effective B field, as explained in appendix A

$$|B_{eff}| = |B|(1 - \frac{v^2}{c^2}) \qquad (14)$$

At variance with what happens in the case of the kinetic terms, this correction cannot be accounted for by a simple replacement of the electron mass with its relativistic counterpart. In fact, even if in the electron magnetic moment the relativistic mass is used, a further correction of $1/\gamma$ in the Pauli term (4c) would still be necessary.

**4. Formalization of the spin-OAM interaction in q-filters**

It is tempting to look for analogies between the effect of the Pauli term $U_p$ in the q-filter, which couples the electron spin and OAM during the propagation, and the well known spin-orbit effect taking place in atoms. In atoms, the spin-orbit coupling originates only as a relativistic effect, as there is no applied magnetic field. The electrostatic potential generated by the nuclei produce, as an effect of relativistic invariance, the well know spin-orbit Hamiltonian, which has the form [24]

$$U_{SO} = \frac{2\mu_B}{\hbar m_e c^2} \frac{1}{r} \frac{\partial V_e}{\partial r} \bar{\sigma} \cdot \bar{l} \qquad (15)$$

where $\mu_B$ is the Bohr magneton and $\bar{l}$ is the OAM vectorial operator. It is easy to verify that

$$U_{SO} \propto [\sigma_z l_z + \sigma_+ l_- + \sigma_- l_+] \qquad (16)$$

where $\sigma_+ = \begin{pmatrix} 0 & 1 \\ 0 & 0 \end{pmatrix}$ and $\sigma_- = \begin{pmatrix} 0 & 0 \\ 1 & 0 \end{pmatrix}$ and $l_z$, $l_+$, $l_-$ are the OAM z component operator and ladder operators, respectively.

If, for example, this potential is applied to a state $|l = 0, \uparrow>$ it is found that

$$< l = 0, \uparrow |(\sigma_+ l_- + \sigma_- l_+)| l = 0, \uparrow> = 0$$

$$< l = 1, \downarrow |(\sigma_+ l_- + \sigma_- l_+)| l = 0, \uparrow> = 1 \qquad (17)$$

This potential hence produces a simultaneous raising or lowering of spin and OAM by one quantum.

Let us now write down the spin-OAM interaction operator resulting from the Pauli term in a quadrupolar magnetic field, where $\bar{B} = B_0/R_0 (y, x, 0)$. Here $B_0$ is the value of the module of the field at a reference radius $R_0$. We thus find

$$U_p \propto \bar{B} \cdot \bar{\sigma} \propto y\sigma_x + x\sigma_y \propto l_+ \sigma_+ + l_- \sigma_- \qquad (18)$$

to prove this relation we used the following paraxial approximation for the angular momentum:

$$l_x \approx p_z y \approx y h/\lambda$$
$$l_y \approx -p_z x \approx -x h/\lambda \tag{19}$$

Which leads, apart from constants, to

$$l_+ \propto x + iy$$
$$l_- \propto -x + iy \tag{20}$$

The formula can be easily extended to higher multipolar geometries, or higher |q| values; E.g. in the case of an hexapole, for which

$$U_p \propto (x^2 - y^2)\sigma_x - 2xy\sigma_y \propto l_+ l_+ \sigma_+ + l_- l_- \sigma_- \tag{21}$$

The similarity of this spin-OAM coupling hamiltonian terms with the spin-orbit atomic ones allows one to develop a formal analogy between our spin-polarization method and the well established polarization scheme based on Mott scattering, in which the polarization is indeed arising from the spin-orbit term. This analogy therefore provides a further argument to explain why our method escapes Bohr's interdiction statement.

**5. Multi-slice algorithm**

The multi-slice simulation algorithm is based on replacing the continuous propagation of the wavefunction along z, under the effect of external fields, with a discrete alternated sequence of thin interaction layers and free-space propagation in given space "slices". In this manner the interaction and propagation can be treated separately, as two simpler problems [ 25].

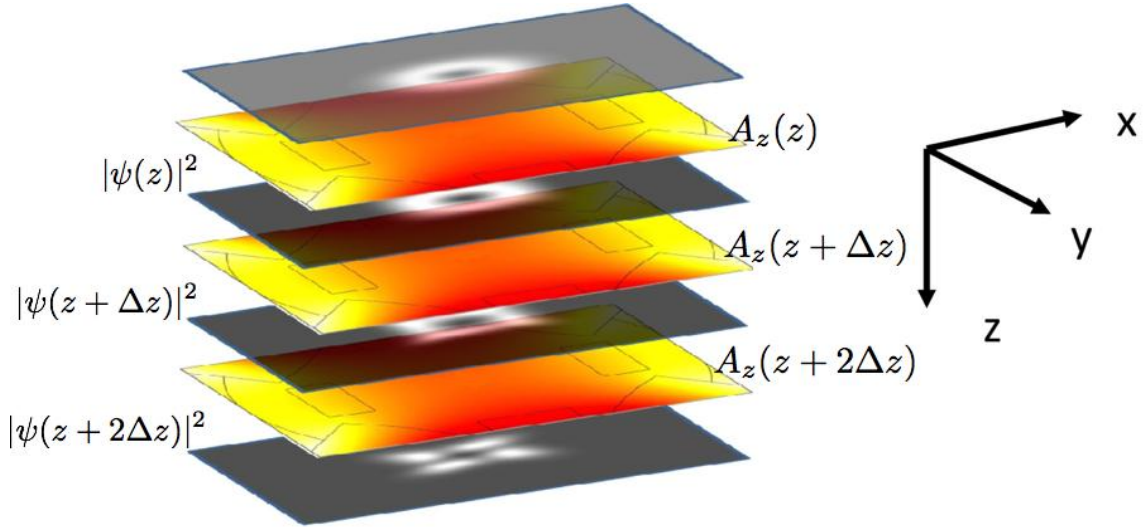

**Figure 1** - Scheme of the algorithm for the multi-slice calculation. After interaction with a small slice of projected potential, a free propagation is calculated between slices. The figure schematizes the wavefunction evolution after the interaction with each potential slice (the $A_z$ component for a quadrupole is also shown).

Explicitly, for the calculation, a 2 component spinor is introduced in the form

$$\psi = \begin{pmatrix} \psi_\uparrow \\ \psi_\downarrow \end{pmatrix} \tag{22}$$

The two wavefunctions appearing in this equation can be treated separately, as long as $H_0$ and $U$ are concerned. The only term mixing them is the Pauli term $U_p$. The normalization condition is $\int \|\psi_\uparrow\|^2 dr + \int \|\psi_\downarrow\|^2 dr = 1$, while the initial amplitude of the two components depends on the polarization state at the entrance of the system. In the case of mixed polarization, the relative phase of the two components should be randomized, to avoid coherent effects. For each single slice i comprised between $z_i$ and $z_{i+1} = z_i + \Delta z$, assuming a B field lying in the xy plane, the effect of the interaction with the fields can be written as [21]

$$\psi(z + \Delta z) = \exp\left[\frac{i}{\hbar v} \int_{z_i}^{z_{i+1}} (U + U_p) \, dz\right] \psi(z) \tag{23}$$

Here we used the velocity v as in eq. 7. This expression assumes implicitly a constant velocity of the electron in the forward direction. Any dependence on the velocity could be included only as higher-order terms.

The interaction term derived from U (eq. 8), excluding the spin, can be rewritten as

$$T(x, y, z_i) = \exp\left(-\frac{ie}{\hbar} \int_{z_i}^{z_{i+1}} A_z(x, y, z) dz + \frac{ie}{\hbar v} \int_{z_i}^{z_{i+1}} V_E(x, y, z) dz + \frac{ie^2}{2m\hbar v} \int_{z_i}^{z_{i+1}} A_z^2(x, y, z) dz\right) \tag{24}$$

However in the case of interest, if $|E|=v|B|$ in every point, the first two terms balance each other and cancel while the third does not and has significant effects on the wavefunction evolution.

If now the Pauli term $U_p$ is introduced, the two spinor component become coupled. The interaction term due to $U_p$ is then written as

$$T_p = \exp\left(\frac{i}{\hbar v} U_p \Delta z\right) \tag{25}$$

using the properties of Pauli matrixes $\sigma_x, \sigma_y$ and assuming a field in the form $(B\cos(\alpha), B\sin(\alpha), 0)$, this can be rewritten as follows:

$$T_p = \begin{pmatrix} \cos\left(2\pi \frac{\Delta z}{\Lambda}\right) & ie^{-i\alpha} \sin\left(2\pi \frac{\Delta z}{\Lambda}\right) \\ ie^{i\alpha} \sin\left(2\pi \frac{\Delta z}{\Lambda}\right) & \cos\left(2\pi \frac{\Delta z}{\Lambda}\right) \end{pmatrix} \tag{26}$$

where $\Lambda = \frac{2\pi \hbar^2 k_0}{m^* \frac{1}{2} g \mu_B B_0}$ and $\alpha$ is taken as $q\theta + \beta$ as in eq. 1. This term produces a mixing of the two spinor components that increases with the slice thickness $\Delta z$.

The multi-slice recipe requires that the interactions with the fields are alternated with free propagations. These can be usually accounted for by the convolution with a propagator function:

$$\psi(z + \Delta z) = P(z) \otimes \psi(z) \tag{27}$$

In the paraxial approximation, the propagator kernel is the following:

$$P(z) = \frac{1}{i\lambda \Delta z} \exp\left[i\frac{\pi}{\lambda \Delta z}(x^2 + y^2)\right] \tag{28}$$

Since the convolution is typically performed in the Fourier space, it is useful to write also its Fourier transform

$$\tilde{P}(K) = \exp\left[i\pi\lambda\Delta z(K_x^2 + K_y^2)\right] \tag{29}$$

where $K_{x,y}$ are the conjugate coordinates of x,y. This expression will be also used for further discussions in appendix D.

The described approach works only for a transverse magnetic field. In the case of a longitudinal field components, there is an additional Hamiltonian term $U = \frac{\hbar e}{m^*} i \overline{A_T} \cdot \overline{\nabla_T}$ (where the T indicates the vector lying in the x,y plane) which is neither a simple phase factor nor a simple free-space propagator, so that the problem becomes more complex. We will consider this term further below, when discussing the effect of the fringing fields.

For the practical implementation of the described method, the Kirkland multi-slice code [25] has been modified in order to add the spinorial representation and the Pauli term. In order to increase the accuracy, a double precision representation of the wavefunction and of the field potentials was used. However the final results were stored on single precision. Finally, for the graphical handling of the simulation, a routine of the STEM CELL software package was added to set the simulation parameters, start the simulations and collect their results [26].

*5.1. Limitations in the multi-slice*

The main numerical limitation in the multi-slice method is in the size of the spatial sampling. In particular, this limits the strength of the magnetic fields that can be accounted for by simulations.

Let us assume for a moment that the propagation effects are negligible. The total phase factor due to the interaction with the magnetic field can be directly calculated by integration of a single uniform potential on the whole lens thickness.

In fact, it is required that the phase shift between neighboring pixels induced by the magnetic field are small, that is

$\Delta \varphi = \varphi_{i+1} - \varphi_i \ll \pi$. We analyze here two cases: if only the $A_z$ term (the third term in eq. 8) matters, one has for example along x

$$\Delta \varphi = -\frac{ie}{\hbar}[A_z(x_{i+1}, y, z) - A_z(x_i, y, z)]\Delta z \tag{30}$$

and therefore to the first order

$$\pi = \Delta\varphi_{Max} = \frac{e}{\hbar} B_y \Delta z \Delta x \tag{31}$$

If the typical sampling is on 1024 pixels for a 100 μm wide slice $\Delta X \approx 100$ nm then $B\Delta z < 2 \cdot 10^{-7}$ T m. This means that fields of 0.001 T can be used only for a thickness of 100 μm. This limits the field and/or the thickness of the optical elements to be simulated.

For a Wien filter, only the $A^2$ term matters and the limit condition becomes

$$\pi = \frac{e^2}{2m\hbar v}|A_z^2(x_{i+1}, x, z) - A_z^2(x_i, y, z)|\Delta z \tag{32}$$

In this case the condition cannot be simply expressed in terms of the magnetic field and thickness of the optical element but the radius of the probe and the shape of the field should also be considered.

We made sure that this condition was satisfied in all considered cases.

**6. Feasibility of the spin-polarization experiment**

Wien filters are electro-optical configurations where the electric and magnetic fields are set up in order to have a compensation between the electrical and magnetic forces. The fields can be constant or space-variant. In the latter case a typical multipolar configuration is used. Wien filters find their application in electron microscopy, for example as monochromators (since the compensation occurs only for a single energy), or in experiments to measure the coherence of the beams [27][28].

The first thing to point out in the description of a Wien filter is that in spite of the fact that we have compensated the field to the first order, the $A^2$ term is not compensated. The semi-classical explanation of this effect (associated with fringe-field effects) was given in the previous article supplementary material [11].

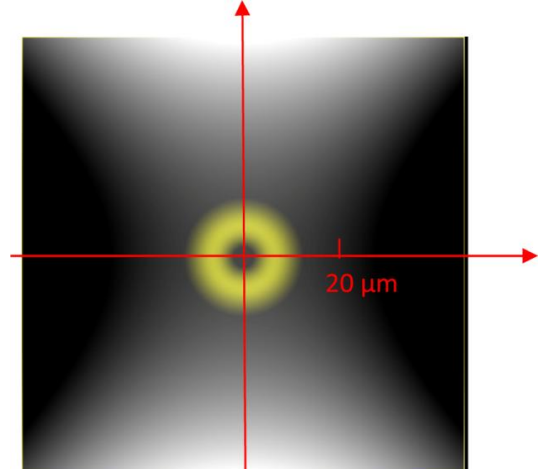

**Figure 2** Spatial distribution of the electrostatic potential, corresponding to the $A_z$ potential of the Wien filter. The doughnut shaped probe before the filter for $\ell=1$ is also shown.

The simulation geometry is shown in figure 2. We assumed as entrance wave a beam in a state with a defined orbital angular momentum per electron OAM=$\ell\hbar$ with $\ell=1$ (which can be obtained, for example, at the exit of a fork hologram) and calculated the evolution at the end of the filter and in the far field, in order to investigate the spatial separation of electrons with different spin.

In more detail, for the entrance wave we used the following spatial dependence of the wavefunction:

$$\psi = N \left(\frac{r}{w}\right)^\ell \cdot \exp\left(-\frac{r^2}{w^2}\right)\exp(i\ell\phi) \qquad (33)$$

Where $N = \sqrt{\frac{2^{\ell+1}}{\pi\ell!w^2}}$ is the appropriate normalization factor. This beam corresponds to a Laguerre-Gauss beam of waist w (maximum intensity at $w' = w\sqrt{\ell/2}$), with the radial number set to 0, and the two spin eigenstates, i.e.

$$\psi_1(z=0) = \Psi|\uparrow\rangle$$
$$\psi_2(z=0) = \Psi|\downarrow\rangle \qquad (34)$$

where $|\uparrow\rangle = \begin{pmatrix}1\\0\end{pmatrix}$ and $|\downarrow\rangle = \begin{pmatrix}0\\1\end{pmatrix}$.

Let us introduce the Rayleigh length $z_R = \frac{\pi w^2}{\lambda}$, which gives the distance at which a Gaussian beam of initial radius w doubles its radius, owing to diffraction. It gives us the order of magnitude of the length scale at which the diffraction effects become relevant. For our simulation, we considered ε=100KeV, a realistic values of w'=10 μm, a slice size of 200 μm, a magnetic field of B=0.003 T at the circle with radius w. The total filter thickness was 5 cm and the number of slices was 20.

Since $z_R$=10 m, while the filter is only few centimeter long, we do not expect a large contribution of diffraction effects inside the device. But the contribution can become larger for higher values of the fields.

The results of the simulation immediately after the filter (near field) are given in figure 3a-h, showing the wavefunction amplitude and phase for different initial spin states and separating the two final spin states.

In this simulation example, the field strength and filter thickness were such that only 0.6% of the beam (as given by the wavefunction integrated absolute square) has undergone spin flip and hence acquired a different OAM value at the exit. The remaining 98.8% is essentially unmodified by the filter.

The exit beam is then let to evolve to its far field by Fourier transforming. The results are shown in figure 3 i-n.

The most evident result is that three states with different radial distributions are well visible in the output, namely corresponding to $\ell$ =0,1,2. Among them, only the state with $\ell$ =1 (the input one) is present in both polarizations: it is hence in a mixed polarization state. The far-field wavefunction patterns show a characteristic "petal" shape, due to the $A^2$ which has a fourfold symmetry.

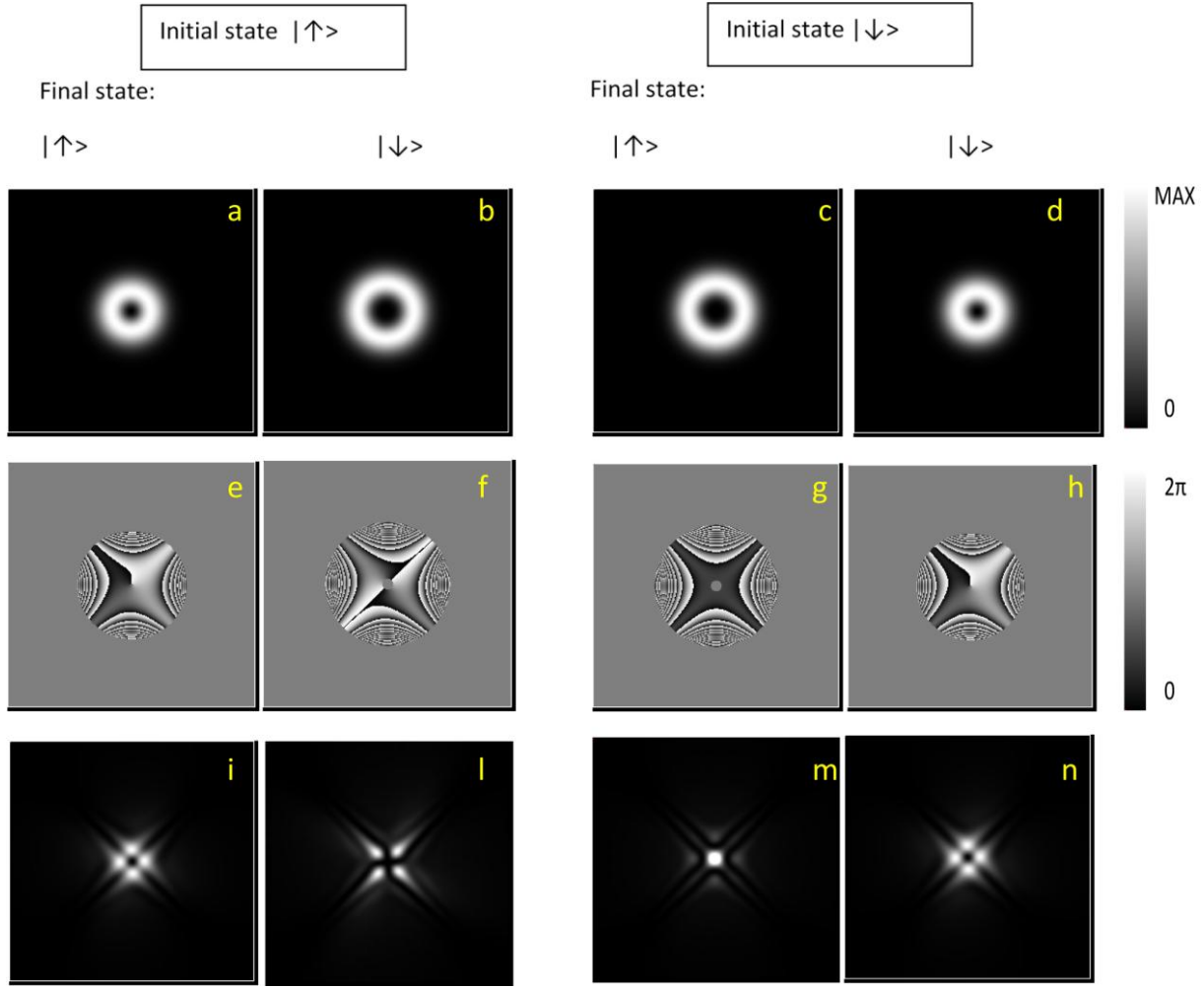

**Figure 3** (a-h) The wavefunction at the exit of the q-filter [(a-d) intensity and (e-h) phase, in gray scale] as a function of the initial and final polarization states. (i-n) Same wave functions after propagation to the far field (only amplitudes).

Due to the different magnetic field strength and filter length, this result cannot be directly compared with the ray tracing calculation of our previous article, but we verified in other cases that the two simulation schemes give comparable results (see appendix B). For the spin-polarization application, the most important outcome here is that only the spin-down input / spin-up output case shows a non-negligible intensity in the beam center (because it has acquired a $\ell$ =0 component). Therefore, by selecting the central region one can filter the spin in an effective way.

*6.2 Expected intensity*

In the precedent article [11] a formula was given for the conversion rate, but it applied to an infinitely thin circle-shaped beam, that is a beam having a single radius. This allowed considering the magnetic field in the formula as a constant, while in a more realistic situation such as that considered here, the field is actually varying across the radial extension of the wavefunction. For this reason, it is easy to demonstrate that a full conversion of the entire beam is actually unreachable. This however does not represent a too severe limitation, as a conversion beyond 50% can still be obtained. The fact that an important, and even dominant, fraction of the electrons remains in the state $\ell =1$ can be considered a practical problem, but not a fundamental limitation. In fact, it can be necessary to reduce the size of the selection pupil with respect to the full conversion case [11], in order to remove the residual transmission of a strong $\ell =1$ component. This produces a further reduction in the filter efficiency. A possible way around this problem could be the use of larger values of |q| and of the OAM of the impinging beam, but this will not be further considered here.

In Table 1, the amount of intensity converted into $\ell=0$ is calculated as a function of the field intensity of the magnetic lenses, for two different values of the electron kinetic energy. The filter length was set to 5 cm and the beam size was 10 µm.

**Table 1.** Fraction of electron spin conversion for different values of the magnetic field in the Wien filter and for two values of the kinetic energy.

| Magnetic field | Kinetic energy of the e-beam | |
| --- | --- | --- |
|  | 100 keV | 40 keV |
| 0.3mT | 6 x $10^{-5}$ | 2.7 x $10^{-4}$ |
| 1mT | 7 x $10^{-4}$ | 0.3 % |
| 3mT | 0.6% | 2.7% |
| 10mT | 6.6% | 25% |

From this table it is clear that the use of lower electron energies makes the spin-polarization easier, by increasing the conversion for the same filter length and magnetic field intensity. As realistic field values at a 10 µm radius are, at the moment, in the range of some mT, a conversion in the order of 1% appears as the most optimistic performance.

In spite of these small numbers, an appropriate aperture could produce a relatively large polarization degree, although at the cost of large intensity reduction. A discussion on the choice of the aperture can be found for example in [29].

*6.3 Coherence issues*

One of the coherence issues arising in Wien filters was studied by Hasselbach [28][30]. It regards the temporal coherence of the electron source which turns eventually in a spatial incoherence effect. In fact, due to the different group velocity of wave packets fractions inside the filter (depending on the local electrostatic potential), two rays experience a different transverse shift of their envelope function even though their phase is practically unchanged. The displacement of wave packets due to the different group velocity can be calculated as

$$\delta z = L \frac{\Delta V}{\varepsilon} \tag{35}$$

Where $\Delta V$ is the difference of potential in two separate optical paths and L is the device length.

Using L=5 cm, $\Delta V$=5 V and $\varepsilon$ = 100KeV, one finds $\delta z$ =2.5 µm this value is comparable with the coherence length of a cold field emission gun and can be therefore a real limiting factor for the size of the applied fields. Since the source of the finite temporal coherence is the energy spread of the source, an equivalent way to see the effect is to introduce a small un-compensation of the Wien filter, namely a small component of the A term proportional to the spread in electron energy. The actual probe profile will then be the incoherent sum of the results obtained for many different energies.

Figure 4 is an example of the propagation where the energy spread is accounted for. Simulations for different energy and consequent unbalancing of the Wien filter have been summed incoherently with a Gaussian weight. Three values of the energy spread have been considered, with Gaussian deviation σ=0.7 eV, 0.3 eV, 0.1 eV, respectively (corresponding to FWHM 1.6eV, 0.7 eV and 0.2eV). The beam energy is 100 keV the filter is 5cm long and with a magnetic field of 0.1 mT. It is clear that with large energy spreads it is impossible to separate the ℓ=0 case.

It is also evident that this factor can be the main limiting factor for the actual conversion rate: with the present sources, this is limited to few percent. A highly monochromatic source could be used to improve this figure, but at the cost of a greatly reduced intensity [31].

Other more conventional issues regarding the spatial and temporal coherence have been already discussed in a precedent paper [32]. It is worth here mentioning that depending on the convergence used in the vortex imaging, the central spot for ℓ =0 can disappear. At the moment this does not seem to be an ultimate limit, though, since many parameters like convergence and the spacing in the fork hologram, or even the kind of hologram to be used can be varied in order to reduce this effect.

*6.4. Fringing Fields*

Since electric and magnetic field forces balance inside the Wien filter, to a first approximation one can assume that their in-plane effects balance also for the fringing fields. In fact, in the region outside the sources the expression for both fields can be calculated starting from the Laplace equation

$$\nabla^2 V = 0 \tag{36}$$

where V represent the electrostatic potential or the magnetic scalar potential. Nevertheless, as in the case of the main field it is necessary to evaluate residual effects due to the Pauli term $U_p$ and $A^2$ potential and an additional term $\overline{A_T} \cdot \overline{\nabla_T}$. For an evaluation of these terms we can base on an analytical model assuming a scaling of the fields with z. Using the scaling function k(z) the expression for the magnetic field and vector potential are provided in appendix C. The simplest expression for these field is obtained for

$$k(z) = \begin{cases} 1 - az & 0 < z < 1/a \\ 0 & z \geq 1/a \end{cases} \tag{37}$$

This scaling can be considered as a first-order approximation. If this simple expression is assumed, the resulting field potential expression is

$$\bar{A} = \frac{B_0}{R_0}\left(\frac{a}{4}xy^2, -\frac{a}{4}x^2y, \frac{1}{2}(1-\mathrm{az})(x^2-y^2)\right) \tag{38}$$

where $B_0$ is the value of the field at a reference distance $R_0$ from optical axis at z=0. The condition $R_0$=w, namely the Laguerre-Gauss radius can be chosen in our case.

The magnetic field is given by

$$\bar{B} = \frac{B_0}{R_0}\left((1-\mathrm{za})y, (1-\mathrm{za})x, -\mathrm{a}xy\right) \tag{39}$$

The derivation and other relevant expressions are given in appendix C. If a->0, this becomes the conventional quadrupolar field expression. The relevant terms are $U_p$ and

$$U = \frac{e^2}{2m^*}A^2 - \frac{\hbar e}{m^*}i\left(A_x\frac{\partial}{\partial x} + A_y\frac{\partial}{\partial y}\right) \tag{40}$$

which is deduced from Eq. (9) by assuming that $V_E$ is exactly compensated in-plane by $A_z$. The effect arising from the second term in Eq. (39) is a deformation of the wavefunction, but this term is typically negligible, as we show in appendix D. The term in $A^2$ provides a weak lensing effect that can be easily taken into account in the simulation, but that does not alter significantly the wavefunction.

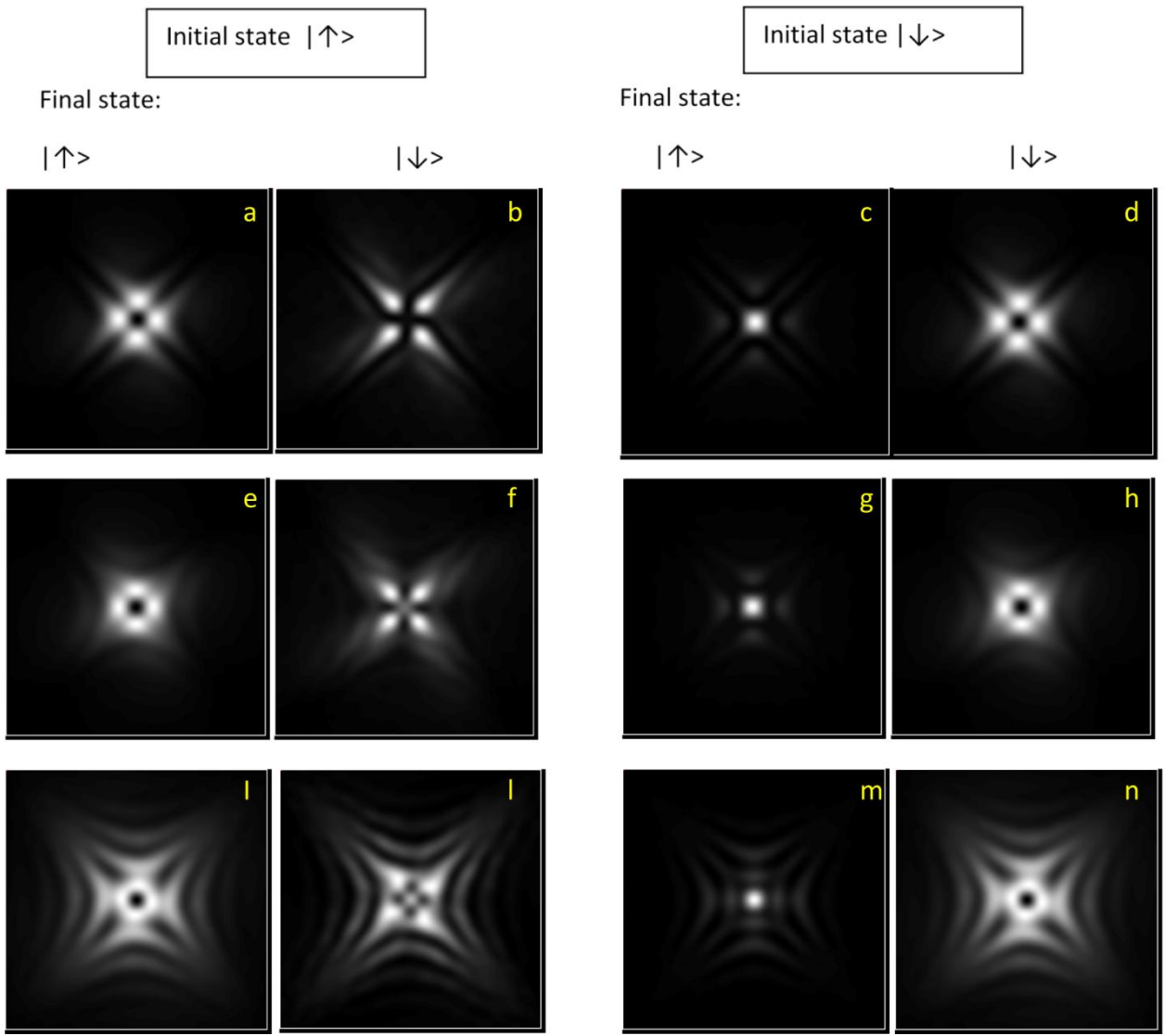

**Figure 4.** The wavefunction intensity at the Fourier (far field) plane after the Wien filter. In this case, the images are obtained as the incoherent average over 5 different energy conditions, weighted by a Gaussian distribution with standard deviation σ = 0.1 eV (a-d), 0.3 eV (e-h) and 0.7 eV (i-n). The effect of the energy spread of the source was accounted for by a residual small miss-compensation of the filter and therefore the appearance of a term in A (the beam radius was w=10 μm and the the field amplitude at 10 μm was 3 mT; the lens was 5 cm thick). It results quite evident that with σ=0.7eV (corresponding to a FWHM=1.5eV) it is not possible to separate the final polarized states.

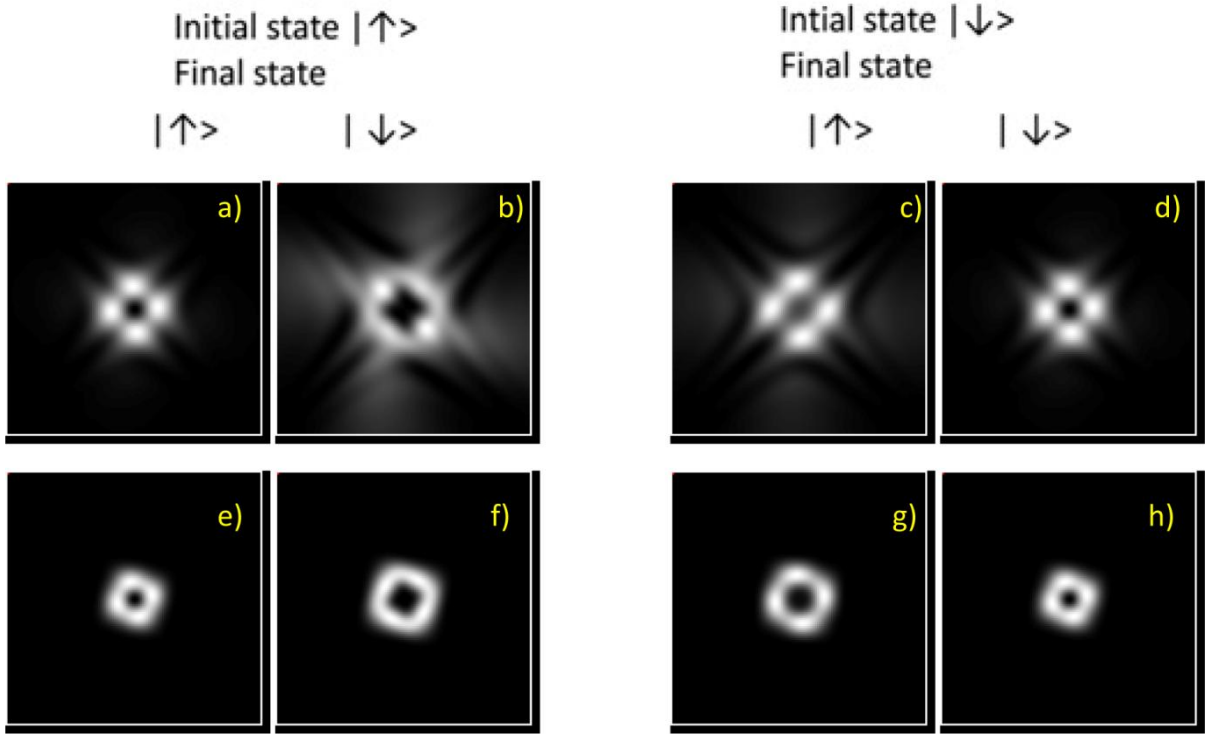

**Figure 5**. Simulation of the effect of a lens made out of the sole fringing fields, without the central region of pure transverse fields, as seen in the Fourier (far field) plane. This geometry could represent also a first approximation of a thin quadupolar lens. Panels a-d refer to the case in which the fringe field regions are $1/a=3$cm long and the field at $r=R_0=10$ μm and $z=0$ (the center of the lens) is 1 mT. Panels e-h refer to the case where $1/a=30$ cm, and the field at $r=R_0=10$μm and $z=0$ was 0.1mT.

The most interesting effect is hence coming from the Pauli interaction. Due to the presence of a complex gradient of the magnetic field, this term couples with the spin in a complicated way, adding a space-variant phase factor (arising from the $B_z$ terms) and spin flip effects (from the transverse components). We find that the most important and disturbing effects occur when the two fringing fields with opposite direction of the field $B_z$ are used.

Figure 5 shows the effect of the sole fringing field (i.e. removing the central Wien-filter region of purely transverse fields) on the wave function in the far field. In figure 5.a-d, the geometry included two regions with fields having opposite direction extending for $1/a=3$cm and the field at $r=R_0=10$μm $z=0$ ($z=0$ is the center of the lens) was set to 1 mT. In figure. 5.e-h, the fringe fields extended for $1/a=30$ cm and the field at $r=R_0=10$μm $z=0$ was set to 0.1 mT. The separation of the spin from these fringe fields alone is clearly not possible, since the state with nominally $\ell =0$ cannot be well separated in the far field

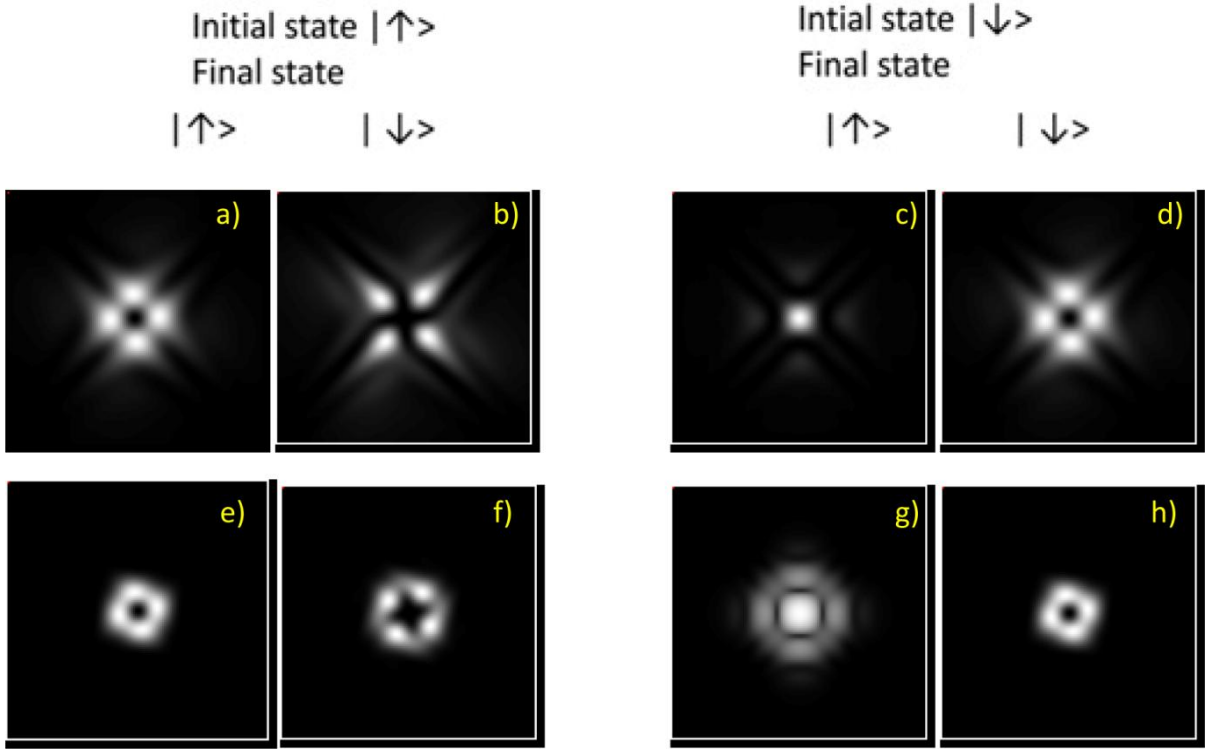

**Figure 6.** Simulation of the effect of the Wien filter inclusive of fringing fields in the Fourier (far field) plane. Panels a-d refer to the case in which the central "stationary" part of the Wien filter is 3 cm long and each fringe field region is $1/a=3$cm long. The field at $r=R_0=10\mu m$ and $z=0$ (the onset of the fringing field) and in the "stationary" part for the same radius is 1 mT. Panels e-h refer to the case in which the central "stationary" part of the Wien filter is 30cm long and each fringing field is $1/a=30$cm long. The field at $r=R_0=10\mu m$ and $z=0$ and in the "stationary" part is 0.1 mT.

Conversely figure 6 shows a simulation for a full Wien filter including a central region of purely transverse fields having the same size as the external regions of fringing fields. The dimension directly correspond to those of figure 5. A net separation of the $\ell=0$ state appears again possible in both considered cases, despite the presence of the fringing fields. This leads us to the conclusion that the fringing fields are not a main limiting factor for achieving the electron spin polarization.

**7. Conclusions**

In this work, we have reanalyzed the feasibility of a realistic electron spin polarizer for applications in electron microscopes. In particular, we have developed a new simulation method based on the multi-slice algorithm to study the spin-wave interaction.

While beam propagation inside the lenses, including fringing fields, does not seem to hamper the possibility of achieving a good spin separation, important limiting issues arise in connection with the achievable field strengths and with finite beam coherence.

Still, there seems to be no fundamental limitation ruling out the feasibility of a spin polarization experiment. The present technological constraints seem to limit the achievable filter efficiency to a few percent, mainly as an effect of chromatic aberrations. Such limitation could be perhaps overcome with the future generation of electron sources.

**Acknowledgments**

One of the authors (VG) would like to thank prof. Pozzi and prof. S. Frabboni for useful discussions. EK, LM and ES acknowledge the support of the FET-Open Program within the 7th Framework Programme of the European Commission, under Grant No. 255914, Phorbitech. E. K. acknowledges the support of the Canada Excellence Research Chairs (CERC) program.

**Appendix A**

This appendix is concerned with the relativistic corrections to be applied to the magnetic field in the spin-field interaction. This effective field accounts for two contributions in the reference system of the laboratory (LS), as both the electric field E and the magnetic field in the Laboratory reference transform to a component of a magnetic field. The formula for the magnetic field B' in the reference system (RS) at rest with the electron is

$$\bar{B}' = \gamma(\bar{B} - \frac{1}{c^2}\bar{v} \wedge \bar{E}) \tag{A1}$$

However in the main part of the compensated Wien filter

$$\bar{E} = -v \wedge \bar{B} \tag{A2}$$

So that

$$\bar{B}' = \gamma\left(B - \frac{v^2}{c^2}B\right) = \frac{1}{\gamma}B \tag{A3}$$

Therefore in the system RS the spin dependent part of the Hamiltonian is

$$U_p = \frac{1}{\gamma}\bar{\mu} \cdot \bar{B} \tag{A4}$$

This result in the reference LS can be also obtained in different way , for example based on Dirac equation which is automatically covariant.

In this context we prefer to describe the evolution matrix due to $U_p$ in RS (see also eq.(23)

$$T = \exp\left[\frac{i}{\hbar}\int_{\tau_i}^{\tau_{i+1}} U_p \, d\tau\right] \tag{A5}$$

The integration limits are calculated in the RS but correspond to the distance $z_{i+1} - z_i$ in the LS through

$$\tau_{i+1} - \tau_i = \frac{1}{\gamma v}(z_{i+1} - z_i) \tag{A6}$$

Therefore the transmission coefficient is

$$T = \exp\left[\frac{i}{\hbar v\gamma}\int_{z_i}^{z_{i+1}} U_p \, dz\right] \tag{A7}$$

This adds an additional coefficient $\frac{1}{\gamma}$ so that the effective B accounting for these effects should be

$$\bar{B}_{eff} = \frac{\bar{B}}{\gamma^2} = \left(1 - \frac{v^2}{c^2}\right)\bar{B} \tag{A8}$$

**Appendix B**

In order to control the reliability of our simulations we compared the propagation predicted by ray-tracing and multi-slice, with identical conditions. For numerical reasons we are limited in the maximum field that we can apply, so we decided to use a smaller radius than those in the previous article. In fact the beam radius w' was reduced to 1 μm and the field applied was 17 and 35, mT at such radius. The lens depth was kept to 4cm. No Pauli interaction was considered in this test.

Fig B1 shows the beam shape for the different values of the magnetic field, showing a very good agreement between the two simulation methods, within the limit in which classical ray tracing can be compared to quantum simulation.

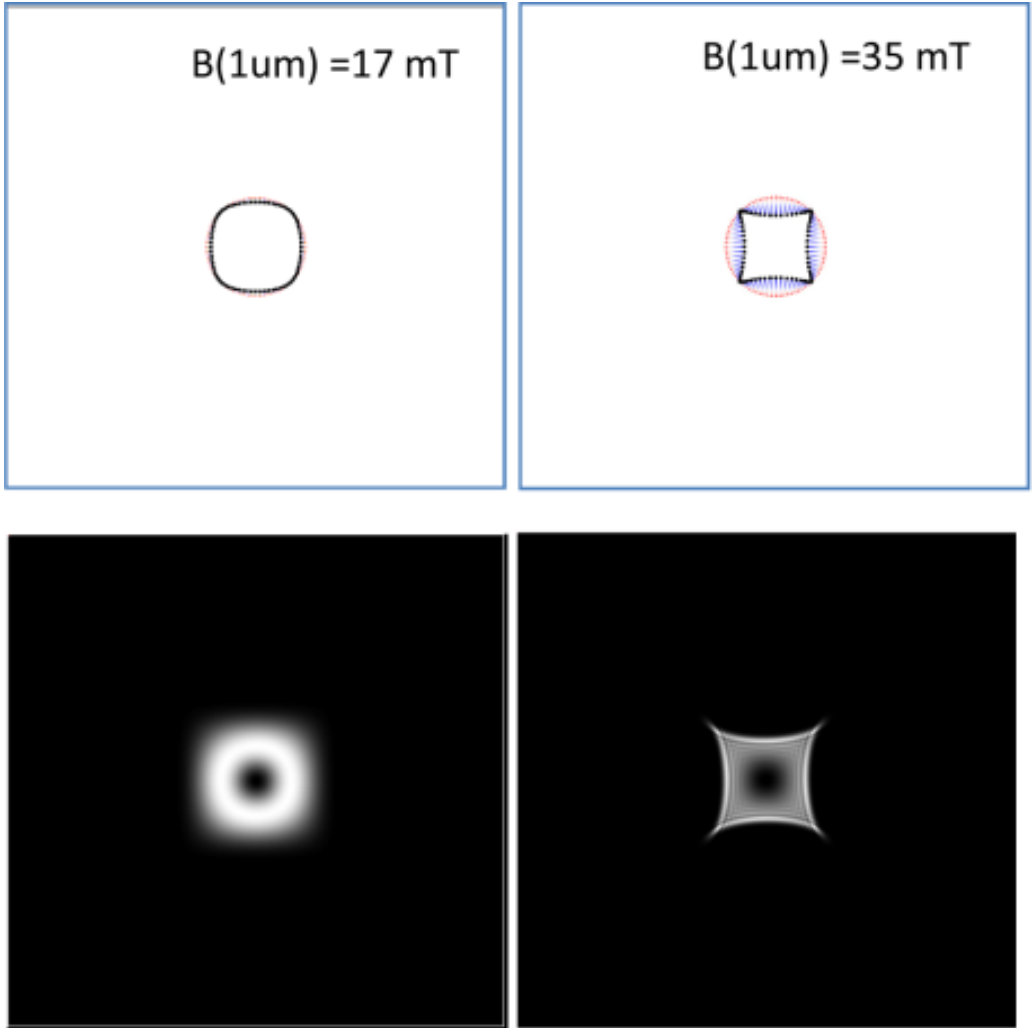

**Figure B1** – Example of comparison between the multi-slice simulation results and the ray-tracing results with the same beam and field parameters. The field intensity is indicated in figure while the beam radius is 1 μm. The lens was 4 cm thick and the beam energy ε was 100 keV.

We notice that while in the quantum simulations based on the multi-slice method the fourfold aberration patterns arise as a consequence of the $A^2$ term in the Hamiltonian, in the classical ray-tracing description the same pattern arises as a consequence of the unavoidable fringe-field effects on the input velocity of the electrons.

**Appendix C**

In order to calculate the fringing field effects we rely on the soft-fringe approximation that leads, for a quadrupole, to the following expression for the electrostatic and magnetic scalar potential [33]

$$V \approx \frac{B_0}{R_0} \frac{k(z)}{2}(x^2 - y^2) \tag{C1}$$

Or with an appropriate axis rotation by 45° that make it consistent with our notation for the magnetic field

$$V \approx \frac{B_0}{R_0} k(z)(xy) \tag{C2}$$

Where k(z) is the scaling factor of the fields with z. Note that this represents an exact solution of the Poisson equation only in the linear case

$$k(z) = 1 - az \tag{C3}$$

Higher order expansion could be possible

$$V \approx \frac{B_0}{R_0} k(z)(xy) - \frac{B_0}{12R_0} \frac{\partial k(z)}{\partial z}(x^3 y + xy^3) + \cdots \tag{C4}$$

That are correct for further polynomial degree of k(z)

Using the first order as an approximate solution the components of the field become

$$B_x = \frac{\partial V}{\partial x} = \frac{B_0}{R_0} k(z) y$$

$$B_y = \frac{\partial V}{\partial y} = \frac{B_0}{R_0} k(z) x$$

$$B_z = \frac{\partial V}{\partial z} = \frac{B_0}{R_0} \frac{\partial k}{\partial z} xy \tag{C5}$$

The vector magnetic potential A that generates this B, compatible with Coulomb gauge and the linear approximation of $K(z)$ is

$$\bar{A} = \frac{B_0}{R_0}\left(-\frac{1}{4}\frac{\partial k}{\partial z}xy^2, \frac{1}{4}\frac{\partial k}{\partial z}x^2 y, \frac{1}{2}k(z)(x^2 - y^2)\right) \tag{C6}$$

For sake of completeness with the 45° rotated system A would have been

$$\bar{A} = \frac{B_0}{R_0}\left(-\frac{1}{8}\frac{\partial K}{\partial z} y^3, \frac{1}{8}\frac{\partial K}{\partial z} y^3, \frac{1}{8} k(z) xy\right)$$

$$\bar{B} = \frac{B_0}{R_0}\left(\frac{1}{2}k(z)x, \frac{1}{2}k(z)y, \frac{1}{2}\frac{\partial k}{\partial z}(x^2 - y^2)\right) \tag{C7}$$

**Appendix D**

We consider here a possible term in the interaction hamiltonian which appears for magnetic field components which are parallel to the z axis. Even if the Wien filter has only transverse fields, such term is present in the fringe fields. It is the term

$$U = \frac{\hbar e}{m^*} i \left(A_x \frac{\partial}{\partial x} + A_y \frac{\partial}{\partial y}\right) \tag{D1}$$

This term cannot be easily included in the multi-slice scheme. The free propagation works as a multiplication by a phase factor in Fourier space, while the spatial potential is a multiplication in real space. Eq. (D1) cannot be dealt with in neither way.

To cope with this term, one possibility is to assume that a limited region is characterized by a constant value of the potential A. In detail, the evolution with this sole term is described in paraxial approximation as

$$\frac{\partial \phi(x, y, z)}{\partial z} = \varsigma(\bar{A}_T \cdot \overline{\nabla}_T)\phi(x, y, z) \tag{D2}$$

where the constant

$$\varsigma = \frac{e\lambda}{2\pi\hbar} \tag{D3}$$

In Fourier space, the evolution of the Fourier Transform of the wavefunction $\tilde{\phi}(K)$

$$\frac{\partial \tilde{\phi}(K,z)}{\partial z} = \varsigma(A_x K_x + A_y K_y)\tilde{\phi}(K,z) \tag{D4}$$

that is solved if A is constant by

$$\tilde{\phi}(K, z+\Delta z) = \tilde{P}'(K)\tilde{\phi}(K,z) \tag{D5}$$

where $\tilde{P}'(K)$ is an additional propagator-like term (see eq. 28 for comparison)

$$\tilde{P}'(k) = \exp\left[i\varsigma(A_x K_x + A_y K_y)\Delta z\right] \tag{D6}$$

This term has the same shape of correction to the propagator with a tilted illumination [25]. In real space, the propagator becomes

$$P'(r) = \delta(\Delta x - \varsigma A_x \Delta z) \otimes \delta(\Delta y - \varsigma A_y \Delta z) \tag{D7}$$

The evolution of wavefunction is therefore

$$\tilde{\phi}(x,y,z+\Delta z) = P'(r) \otimes \tilde{\phi}(x,y,z) \tag{D8}$$

Therefore, the wavefunction is translated, in the transverse plane, by a length proportional to $\overline{A_T}\Delta z$. The additional step now is to assume that $A_x$ and $A_y$ vary slowly with position. In such case, the translation due to the local $\overline{A_T}$ induced tilt is position dependent. Eq. 6 is locally valid as infinitesimal translation, while eq. 5 is not.

This approximation can be demonstrated to work for example when a fixed field B is along z and A is

$$\overline{A} = \frac{B}{2}(-y, x, 0) \tag{D9}$$

This implies a translation in both x and y

$$\begin{aligned} x' &= x + \tfrac{1}{2}\varsigma B y \Delta z \\ y' &= y - \tfrac{1}{2}\varsigma B x \Delta z \end{aligned} \tag{D10}$$

which is an infinitesimal rotation. If this evolution is performed 2n times, one obtains

$$\begin{aligned} x' &= x \cdot \cos(n\varsigma B \Delta z) - y \cdot \sin(n\varsigma B \Delta z) \\ y' &= y \cdot \cos(n\varsigma B \Delta z) + x \cdot \sin(n\varsigma B \Delta z) \end{aligned} \tag{D11}$$

which is the exact evolution in a constant field along z.

In the case of the fringing field the shape of the field is more complex and for generality we applied the translation through a numerical approach where both real and imaginary part are interpolated with sub-pixel precision by spline algorithms in order to perform the appropriate rotation.

In the specific case it turns out that within the linear approximation for the fringes shape the total translation (assuming it only occurs in one step) does not depend on the fringe width.

We can calculate the size of the displacement using

$$A_x = \frac{B_0}{R_0}\left(\frac{a}{4}xy^2\right) \tag{D12}$$

For example when x=y=$R_0$ a=1/$\Delta z$

$$A_x = A_y = \frac{B_0}{R_0}\left(\frac{1}{4\Delta z}R_0^3)\right) \tag{D13}$$

we find

$$\Delta x = \varsigma A_x \Delta z = 0.5\ \varsigma B_0 R_0^2 \tag{D14}$$

The displacement does not depend on the size of the fringing field but on the lateral distance from the optical axis. For a B=1mT $R_0$=0.1 mm the displacement is of the order of 10nm, namely below the typical size of a pixel. In most cases this term can be therefore neglected. Nevertheless we accounted for this term in all fringing field simulations and verified it had no sizable effect.